\documentclass[aps,prb,twocolumn,superscriptaddress,showkeys]{revtex4}
\usepackage{bm,amsmath,graphicx}
\begin{document}

\title{On diffraction and absorption of X-rays in perfect crystals}

\author{S\'ergio L. Morelh\~ao}
\email{morelhao@if.usp.br}
\affiliation{Instituto de F\'{\i}sica, Universidade de S\~ao Paulo, CP 66318, 05315-970 S\~aoPaulo, SP, Brazil}
\author{Luis H. Avanci}
\email{lhavanci@if.usp.br}
\affiliation{Instituto de F\'{\i}sica, Universidade de S\~ao Paulo, CP 66318, 05315-970 S\~aoPaulo, SP, Brazil}

\date{\today}

\begin{abstract} 
This work reports the incompatibility of the Stokes relation for time reversibility of wave's propagation and the usage of complex atomic form-factor in diffraction theories. The implications behind a diffraction theory that obeys the Stokes relation is discussed.
\end{abstract}

\keywords{Diffraction physics, Anomalous dispersion, Standing waves, Foundations of X-ray crystallography}

\maketitle

\section{Introduction}

Interaction properties of X-rays with atoms are describable by complex atomic scattering factors in the widely used format 

\begin{equation}
f = f_0+f'+if''.
\label{casfeqn}
\end{equation}
Atomic resonance theories \cite{jame1948, born1980, crom1965, crom1970, kiss1995} provide the physical bases for the $f'$ and $f''$ corrections of $f_0$, the scattering factor for incident photons with energies away from the absorption edges of the atom.

In an aggregate of atoms where the scattering amplitudes of the individual atoms have random phases, the material's index of refraction is given by

\begin{equation}
n_r=1-\frac{r_e \lambda^2}{2\pi V}\sum_n (f_0+f^{\prime}+if^{\prime\prime})_n.
\label{iofreqn}
\end{equation}
$n$ runs over all atoms in the volume $V$, $\lambda$ is the X-ray wavelength, and $r_e=2.818\times10^{-5}$\AA~is the classical electron radius.

$(1/V)\sum_n(f_0+f')_n$ provides the effective electron density, and it is responsible for the difference from unit of the real part of the index of refraction, and then, for the effect of refraction. The imaginary part of $n_r$, proportional to $\sum_n f''_n$, is responsible for the intensity attenuation of the travelling X-ray waves in the medium. 

In crystalline materials, the scattering and absorption properties of the atoms are summarized in the structure factor

\begin{eqnarray}
{\cal F}_H = \sum_n(f_0+f'+if'')_n \exp(+2\pi i{\bm H}\cdot{\rm r}_n) \nonumber \\ 
           = |{\cal F}_H|e^{i\alpha_H},
\label{ssfeqn}
\end{eqnarray}
which is obtained as the scattering amplitude by the crystal's unit cell for a given reflection H, whose diffraction vector is ${\bm H}$. Atomic positions in the unit cell are specified by ${\rm r}_n$.
 
The above structure factor expression, Eq.~(\ref{ssfeqn}), is well-accepted since it is able to provide an explanation for different intensities of the H and $\bar{\rm H}$ reflections, \cite{cost1930, geib1932, bijv1951} i.e. for the breaking of the Friedel's Law. \cite{mgfr1913} It is possible because in noncentrosymmetric crystals $|{\cal F}_H|\neq|{\cal F}_{\bar{H}}|$ owing to the $f''_n$ scattering amplitudes, commonly called anomalous or resonant scattering amplitudes. 

There are several structure phasing methods currently in use based on the anomalous scattering phenomenon, such as multi-wavelength anomalous dispersion or isomorphous replacements of anomalous scatters (see for reviews Refs. \onlinecite{cati1981}, \onlinecite{giac1999}, and \onlinecite{hell1992}). These methods have allowed X-ray crystallographers to solve thousands of protein structures, which is a strong evidence that complex atomic scattering factors properly describe the interaction physics of the X-ray waves with the atoms. Moreover, near absorption edges in cases where $f_0+f'=0$, the dynamical diffraction of only the resonant scattering term $if''$ has been theoretically predicted and demonstrated experimentally. \cite{kato1992,fuka1993, negi2004}

The initial motivation for this work was related to energy conservation in approximated solutions of multi-beam scattering problems, \cite{more2005a} when applying the Stokes relation of thin film optics \cite{knit1976} to each lattice plane of a single crystal. It has led to an incompatibility with the usage of complex atomic scattering factors, e.g. Eq.~(\ref{casfeqn}), in diffraction theories that exists even in two-beam diffraction cases, as further discussed here.

An apparent incompatibility arrives because the reflection and transmission coefficients derived from Stokes relation do not allow absorption to be treated as anomalous-resonant scattering amplitudes. But, on the other hand, these coefficients preserve the sum of probabilities for reflection, transmission and absorption of photons at each lattice plane, and they provide an accurate description of the X-ray diffraction in perfect crystals with thickness varying since the kinematical regime of diffraction (few lattice planes) to the dynamical one (very thick crystals). Then, the point of conflict relies on how these coefficients that provide a good description of the diffraction phenomenon can not be in agreement with atomic resonant theories.

To properly explain the implications behind a diffraction theory that obeys the Stokes relation and to discuss its consequence on X-ray crystallography, this article has been divided into three Sections (II, III, and IV):

\medskip {\bf Section II}: Derivation of reflection and transmission coefficients from the Stokes relation for a general quantum particle. Recursive equations for calculating the reflectivity of absorbing crystals with a finite number of lattice planes are also obtained. These recursive equations are extremely simple and, even thus, capable to describe important features of the diffraction phenomenon such as kinematical diffraction, primary extinction, intrinsic width of Bragg reflections in either kinematical or dynamical diffraction regimes, and phase shift of the diffracted waves across the reflection domain.

\medskip {\bf Section III}: Application of the recursive equations, derived in Section II, to the particular case of X-ray diffraction. It emphasizes the formation of standing waves as necessary to explain the intensity differences of Friedel reflections in noncentrosymmetric crystals when the structure factor is calculated without the resonant term $if''$.

For a given example, the anomalous signal (intensity difference from a pair of Friedel reflections) calculated here as an effect of absorption modulation by standing waves \cite{laue1949, borr1954, batt1962, batt1964a} is compared to that obtained by the usual expression of the structure factor, as well as to experimental data. This comparison allows a clear distinction between absorption and interference effects owing to the $if''$ term. It also demonstrates that the majority of the anomalous signal amplitude has its source on absorption modulation and therefore it should be a function of the thickness when the crystal is very small (below the primary extinction length). Implication on the accuracy of phasing methods based on anomalous dispersion is discussed. 

\medskip {\bf Section IV}: A discussion is given on a possible physical explanation for the apparent incompatibility of the Stokes relation [e.g. Eq.~(\ref{bse1})] and the usage of the anomalous-resonant term in the expression of the structure factor, i.e. the usage of $if''$ in Eq.~(\ref{ssfeqn}). The central point of discussion relays on causality, collapse of quantum probability amplitudes for photons, and stimulated photon emission by excited atoms.

\section{Stokes relation in diffraction theories}

The elastic scattering of quantum particles (photons, electrons, neutrons, ...) by a thin plane of matter can be described by introducing reflection $R$, and transmission $T$, coefficients so that $\psi_R=R\psi_0$ and $\psi_T=T\psi_0$ are the probability amplitudes (or the vectorial electric field amplitudes in the case of photons) for the reflected and transmitted particles, respectively. When the probability amplitude $\psi_0$ of the incident particles is uniform over a significant area of the plane, regarding the dimensions of its building blocks, the particles are either specular reflected or transmitted along the same direction of the incident ones, as for source S$_1$ in Fig.~1. $|R|^2$ and $|T|^2$ are therefore the reflection and transmission probabilities of detecting the particles along these directions, as indicated in Fig.~1 by the detectors D$_1$ or D$_2$, respectively. For non-absorbing and non-dispersive planes $|R|^2+|T|^2=1$, otherwise $a=1-|R|^2-|T|^2$ is the effective absorption probability accounting for the probability of the particles to be in fact absorbed (photoelectron absorption) or inelastic scattered at random directions. 

\begin{figure}
\includegraphics[width=3.2in]{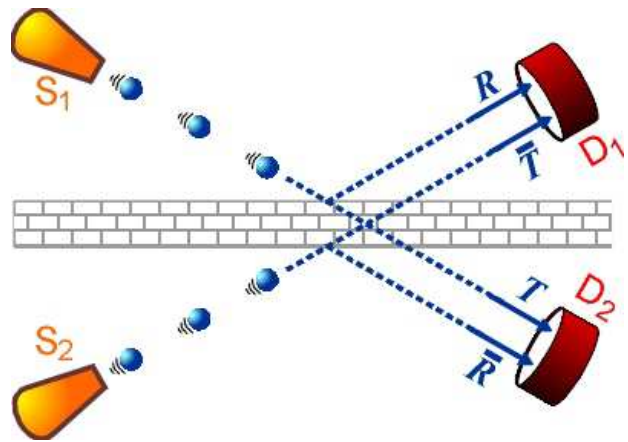}
\caption{Specular scattering of quantum particles (photon, electrons, neutrons, ...) from identical sources S$_1$ and S$_2$, symmetrically displaced at both side of a thin plane of matter. The reflected and transmitted probability amplitudes are given by $\psi_R=R\psi_0$, $\psi_T=T\psi_0$, $\psi_{\bar{R}}=\bar{R}\psi_0$, and $\psi_{\bar{T}}=\bar{T}\psi_0$.}
\end{figure}

The reflection and transmission coefficients are in principle different, given by $\bar{R}$ and $\bar{T}$, for incidence on the other side of the plane, as for source S$_2$ in Fig.~1. Since the sources are identical and symmetrically displaced on each side of the plane, the sum of the counting rates from both detectors are equal to the total emission rate of the sources. It means that, for non-absorbing planes, $|\psi_R+\psi_{\bar{T}}|^2+|\psi_{\bar{R}}+\psi_T|^2=2|\psi_0|^2$, and hence

\begin{equation}
R\bar{T}^{*}+R^{*}\bar{T}+\bar{R}T^{*}+\bar{R}^{*}T=0.
\label{bse1}
\end{equation}
It stipulates some phase relationships between the reflected and transmitted amplitudes, as usually obtained for laser beam splitters \cite{loud2000} or, equivalently, by the Stokes relation for time reversibility of wave's propagation. \cite{knit1976}.

Without loosing generality, the probability amplitude coefficients can be written as

\begin{eqnarray}
R&=&\pm i|R|\>e^{i(\delta+\bar{\varphi})},~~T=|T|\>e^{i\varphi},\nonumber\\
\bar{R}&=&\pm i|\bar{R}|\>e^{i(\bar{\delta}+\varphi)},~~{\rm and}~~\bar{T}=|\bar{T}|\>e^{i\bar{\varphi}}
\label{phaeqn}
\end{eqnarray}
where according to Eq.~(\ref{bse1}), 

\begin{equation}
|R||\bar{T}|\sin\delta+|\bar{R}||T|\sin\bar{\delta}=0. 
\label{bse2}
\end{equation}

$\varphi$ and $\bar{\varphi}$ are the phase delays across the plane, whose thickness $d$ is comparable to the particle's wavelength $\lambda$. Since the incidence angle $\theta$ is the same on both sides

\begin{equation}
\varphi = \bar{\varphi}=-\>\>\frac{2\pi}{\lambda}d\sin\theta\>.
\label{phdeqn}
\end{equation}

$\delta\pm90^{\circ}$ and $\bar{\delta}\pm90^{\circ}$ are the amount by which the phases of the reflected particles differ from the phases of the transmitted ones. If $\delta=\bar{\delta}=0$, the phase of the reflected particles are $90^{\circ}$ shifted with respect to the transmitted ones. \cite{zach1945, jame1948, warr1969} 

\subsection{Reflection and transmission coefficients for N-plane crystals}

In crystals, each individual scattering plane stands for an element of periodicity composed, in general, of several atomic layers that repeat themselves with period $d$ in order to build up the crystal. The scattering properties of each element of periodicity (lattice plane), when specified by reflection and transmission coefficients, Eq.~(\ref{phaeqn}), allow to calculated the $R_N$, $T_N$, $\bar{R}_N$, and $\bar{T}_N$ coefficients for crystals with $N$ planes in reflection geometry. 

To carry out this calculation, let first analyze the two-plane case. In such case, the particles can suffer several reflections in-between the planes before exiting, as depicted in Fig.~2. However, even when there is no empty space $h$ between the planes ($h\rightarrow0$), the contributions of the several consecutive reflections have to be taken into account. It provides that for $N=2$

\begin{subequations}
\begin{equation}
R_2=R+TR\bar{T}+TR\bar{R}R\bar{T}+...=R+TR\bar{T}\sum_{n=0}^{\infty}(\bar{R}R)^n
\label{r2eqn}
\end{equation}
and 
\begin{equation}
T_2=TT+TR\bar{R}T+TR\bar{R}R\bar{R}T+...=TT\sum_{n=0}^{\infty}(R\bar{R})^n
\label{t2eqn}
\end{equation}
\label{r2t2eqn}
\end{subequations}
for incidence from the top-left, as in Fig.~2. In the other symmetrical situation, which corresponds to incidence from the bottom-left, e.g. source S$_2$ in Fig.~1, the $\bar{R}_2$ and $\bar{T}_2$ coefficients are obtained by analogous procedure. Note that these coefficients are very similar to those in the Airy's formula of the Fabry-Perot interferometer, and that their expressions can be simplified by using $\sum_{n=0}^{\infty}z^n=1/(1-z)$ for $|z|<1$.

Although it would be possible to add plane-by-plane with the above procedure, a fast recursive formula to go from very thin to semi-infinity crystals (of infinity thickness) is obtained by building the crystals in geometrical progression of $N=2^n$ planes ($n = 1, 2,...\>$). Since the equations to calculate the coefficients of $N$ planes from those coefficients of $N/2$ planes are identical to Eqs.~(\ref{r2t2eqn}), 

\begin{subequations}
\begin{equation}
\left[\begin{array}{c} 
R_N\\ 
\bar{R}_N
\end{array}\right]=
\left(1+\frac{T_{N/2}\bar{T}_{N/2}}{1-\bar{R}_{N/2}R_{N/2}}\right)\left[\begin{array}{c}
R_{N/2}\\
\bar{R}_{N/2}
\end{array}\right] 
\label{rneqn}
\end{equation}
and
\begin{equation}
\left[\begin{array}{c} 
T_N\\ 
\bar{T}_N
\end{array}\right]= \frac{1}{1-R_{N/2}\bar{R}_{N/2}}\left[\begin{array}{c}
T_{N/2}^2\\
\bar{T}_{N/2}^2
\end{array}\right] 
\label{tneqn}
\end{equation}
\label{rntneqn}
\end{subequations}
are the amplitude coefficients for $N$-planes in a crystal of thickness $Nd$. 

\begin{figure}
\includegraphics[width=3.2in]{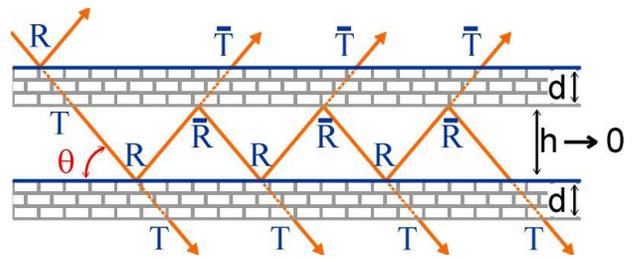}
\caption{Reflection and transmission coefficients of two planes. Several consecutive reflections occur in-between the planes, as in the resonant cavity of a Fabry-Perot interferometer. They have to be taken into account even when there is no empty space $h$, between the planes, i.e. when $h\rightarrow0$. $R$, $T$, $\bar{R}$, and $\bar{T}$ are the amplitude coefficients of each individual plane, as given in Eq.~(\ref{phaeqn}).}
\end{figure}

\subsection{Absorption}

The absorption probability 

\begin{equation}
a=\mu d/\sin\theta
\label{apeqn}
\end{equation}
is given in terms of the linear absorption coefficient $\mu$ of the material. It does reduce the transmission probability per lattice plane according to

\begin{equation}
|T|^2=(1-|R|^2)\>e^{-a}\simeq(1-|R|^2)(1-a)\simeq1-|R|^2-a 
\label{tpeqn}
\end{equation}
since $a|R|^2 \ll |R|^2+a \ll 1$. 

\subsection{Intrinsic width and phase shift in rocking curves}

As a function of the incidence angle $\theta$ ~\textemdash~ the variable angle in rocking curves ~\textemdash~ the reflection coefficient of N planes, 

\begin{equation}
R_N(\theta)=|R_N(\theta)|\>e^{i\Psi(\theta)}
\label{refleqn}
\end{equation}
determines the intrinsic profile of the diffraction peak, i.e. the intensity reflectivity curve $|R_N(\theta)|^2$, as well as the phase

\begin{equation}
\Psi(\theta)=\Omega(\theta)+\delta,
\label{phaseqn}
\end{equation}
of the reflected particles with respect to the incident ones. $\Omega(\theta)$ stands for the phase dependence with the incidence angle, and $\delta$ is determined by the internal structure of the lattice planes.

The center of the reflectivity curve is observed at the Bragg angle $\theta_B$. At this angle, the phase of $T\bar{T}$ in Eq.~(\ref{r2eqn}) is an integer number of $2\pi$, i.e. $2\varphi=-2m\pi$ ($m=\pm1,\>\pm2,...$) and hence, for the first order reflection ($m=1$) we have that $\sin\theta_B = \lambda/2d$.

A few examples on the behavior of the intensity reflectivity curve $|R_N(\theta)|^2$ as a function of crystal thickness, reflection and absorption probabilities are shown in Fig.~3. The phase shift $\Omega(\theta)$ across the total reflection domain of the rocking curve is also shown (inset of Fig.~3). In all calculated curves, $|R|=|\bar{R}|$ and $|T|=|\bar{T}|$.

Absorption is observed to symmetrically round (regarding the center of the peak) the flat-top of the total reflection domain, and to reduce both the integrated intensity (area of the rocking curve) as well as its maximum value.

The analysis of the reflectivity profiles for thick crystals leads to an empirical equation for the full width of the half maximum (FWHM),

\begin{equation}
W = \frac{2}{3}\sqrt{|R||\bar{R}|}\tan\theta_B.
\label{iweqn}
\end{equation}
For a given Bragg angle, the width of the rocking curve is essentially determined by the particles specular reflection probability $|R|^2$, at each lattice plane. When the interaction physics of the particles with the planes are known, the FWHM can be calculated by means of more specific parameters such as electronic density, atomic numbers (neutron scattering), X-ray polarization, particle energy, {\em etc}. On the other hand, the reflection probability can be estimated only from the values of $W$ and $\theta_B$. For instance, the silicon 111 reflection at about 8KeV presents $W\simeq10''$ and an incidence angle close to $14^{\circ}$, providing $|R|^2\simeq8.5\times10^{-8}$.

\begin{figure}
\includegraphics[width=3.2in]{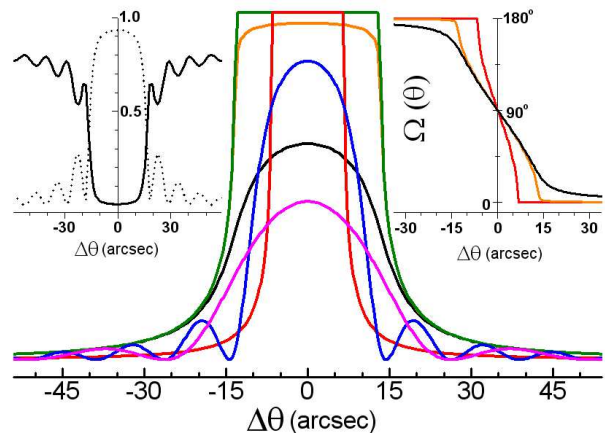}
\caption{Intensity reflectivity curves $|R_N(\theta)|^2$ in crystals of thickness $Nd$, according to Eqs.~(\ref{rntneqn}). $\Delta\theta=\theta-\theta_B$, $\lambda=1.54$\AA, and $d=3.14$\AA~in all of the following cases: (1) [$N$, $|R|^2$, $\mu (cm^{-1})$] = [2048, $16\times10^{-8}$, $0.01$] (pink curve); (2) [4096, $16\times10^{-8}$, $0.01$] (blue curve); (3) [$\infty$, $16\times10^{-8}$, $0.01$] (red curve); (4) [$\infty$, $64\times10^{-8}$, $0.01$] (green curve); (5) [$\infty$, $64\times10^{-8}$, $200$] (orange curve); (6) [$\infty$, $64\times10^{-8}$, $8000$] (black curve); and (7) [4096, $64\times10^{-8}$, $400$], for which both $|R_N(\theta)|^2$ (dashed curve) and $|T_N(\theta)|^2$ (black-line curve) are shown at the top-left inset. For cases (3), (4), (5) and (6), the phases of $R_N(\theta)$ are given in the top-right inset; they correspond to $\Omega(\theta)$ since $\delta=0$ in Eq.~(\ref{phaseqn}) was used in all simulations.}
\end{figure}

\section{X-ray diffraction}

In the case of X-ray diffraction, the $R$ and $\bar{R}$ reflection coefficients of a single lattice plane should not bear the resonant term $if''$. Otherwise, in general, the relationship in Eq.~(\ref{bse2}) will not be fulfilled in crystals without a center of symmetry (see Section IV for more details). To accomplish the condition imposed by this relationship, the scattering and absorption of X-rays by individual atoms are treated here as follow.

The atomic scattering amplitude taking place in interference phenomena is given by  
\begin{equation}
\sigma_S(2\theta) = r_e \lambda(f_0+f^{\prime})|C|,
\label{scseqn}
\end{equation}
while the atomic absorption cross section 
\begin{equation}
\sigma_A = 2 r_e \lambda f^{\prime\prime}
\label{acseqn}
\end{equation}
accounts for photoabsorption. $|C|=1$ or $\cos2\theta$ stands for the $\sigma$ or $\pi$ polarization components, respectively. $f_0$ is the main scattering factor through an angle $2\theta$ from the incident beam direction. More details on $f_0$, $f'$, and $f''$ values are available on the {\em International Tables for Crystallography}. \cite{inte2001}

A single layer of atoms, all of the same kind, with $M$ atoms per unit area scatters an incident plane wave of amplitude $E_0$ according to \cite{zach1945, warr1969}

\begin{equation} 
E_{SL} = -i\frac{M}{\sin\theta}\sigma_S(2\theta) E_0 = R_{SL} E_0. 
\label{slaeqn}
\end{equation} 
$|R_{SL}|^2$ is the probability to measure the specular scattering of photons by this single layer of atoms, while the absorption probability is given by 

\begin{equation} 
a_{SL} = \frac{M}{\sin\theta}\sigma_A. 
\label{asleqn}
\end{equation} 

To generalize the reflection coefficient $R_{SL}$, from the single layer to elements of periodicity composed of several atomic layers, as illustrated in Fig.~4, the number of atoms per unit area is written in terms of the crystal's unit cell, so that

\begin{equation}
R_{SL}=-i\frac{M}{\sin\theta} \sigma_S \rightarrow
R=-i\frac{d}{V_c\sin\theta}\sum_{n}\sigma_S(n)e^{-2\pi i\bm{H}\cdot\bm{r}_n}.
\label{mtof}
\end{equation}
$n$ runs over all atoms in the unit cell, whose positions and scattering amplitudes are given by $\bm{r}_n$ and $\sigma_S(n)$, respectively. $\bm{H}$ is the diffraction vector of the chosen reflection H. $V_c$ is the unit cell volume and $d = 1/|\bm{H}|$ as usual. The exponents, or geometrical phase factors, are necessary to account for the phases of the photons coherently scattered at the atomic sites, and its `` - '' (minus) signal will be justified later on.

By replacing Eq.~(\ref{scseqn}) into Eq.~(\ref{mtof}) and defining 

\begin{equation}
F_H = \sum_n (f + f^{\prime})_n\exp(-2\pi i{\bm H}\cdot{\bm r}_n)=|F_H|e^{i\delta_H}
\label{sfeqn}
\end{equation}
as the structure factor of reflection H, the reflection coefficients for individual lattice planes are 

\begin{equation}
\left[\begin{array}{c} 
R\\ 
\bar{R}
\end{array}\right]=-i\frac{r_e\lambda|C|d}{V_c\sin\theta}\left[\begin{array}{c}
F_{\rm H}e^{i\varphi}\\
F_{\bar{\rm H}}e^{i\bar{\varphi}}
\end{array}\right] 
\label{rlpeqn}
\end{equation}
where $F_{\bar{\rm H}}$ stands for the structure factor of reflection $\bar{\rm H}$ with diffraction vector ${-\bm H}$. Temperature factors are implicit in the $f$ and $f'$ values. The photoabsorption probability by the lattice plane is obtained with the same procedure, i.e.

\begin{equation}
a_{SL}=\frac{M}{\sin\theta} \sigma_A \rightarrow a=\frac{d}{V_c\sin\theta}\sum_{n}\sigma_A(n).
\label{alpeqn}
\end{equation}
No geometrical phase factor is necessary since $\sigma_A(n)$ is not a scattering amplitude.

To make certain that the above results are correct, they can be compared with those already known from X-ray diffraction theories. The width of the total reflection domain, characterized by top-hat shape of the reflectivity curves in Fig.~3, also known as the Darwin width, \cite{darw1914a, darw1914b, prin1930} is obtained by replacing Eq.~(\ref{rlpeqn}) into Eq.~(\ref{iweqn}), and then, the FWHM is given in terms of the same parameters used to calculate the Darwin width $W_D$, i.e.

\begin{equation}
W=\frac{2d}{3}\frac{r_e\lambda|C|(F_{\rm H} F_{\bar{\rm H}})^{1/2}}{V_c
\cos\theta}=\frac{\pi}{3}W_D,
\label{dweqn}
\end{equation}
see Ref. ~\onlinecite{auth1998} for a comparable expression of $W_D$.
 
The phase shift of $180^{\circ}$ across the rocking curve, $\Omega(\theta)$ in the inset of Fig.~3, agrees with that calculated by the dynamical diffraction theory, which is known as the dynamical phase shift \cite{auth1986, auth2001}. Only the exact center of the reflectivity curves, at $\theta_B$, are slightly different since $\lambda$ has not been corrected for the effect of refraction.

Eqs.~(\ref{apeqn}), (\ref{acseqn}) and (\ref{alpeqn}) can be combined to provide the linear absorption coefficient as 

\begin{equation}
\mu = \frac{1}{V_c}\sum_n \sigma_A(n)=\frac{2r_e\lambda}{V_c}\sum_n
f_n^{\prime\prime},
\label{laceqn}
\end{equation}
which is the usual expression for calculating $\mu$ of a crystal in terms of its unit cell, see for instance Ref. ~\onlinecite{batt1964a}. 

\begin{figure}
\includegraphics[width=3.2in]{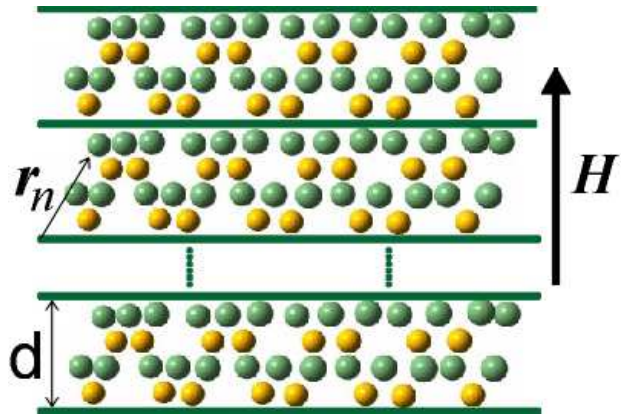}
\caption{Each crystal reflection stands for an element of periodicity composed in general of several atomic layers. Regardless the chosen origin, the ${\bm r}_n$ atomic positions in the elements of periodicity (lattice planes) are specified from bottom to top, in the same sense of the diffraction vector ${\bm H}$. This is the usual convention in X-ray crystallography.}
\end{figure}

\subsection{Absorption modulation by standing waves}

In order to fulfill the condition imposed by the Stoke relation, e.g. Eq.~(\ref{lasteqn}), structure factors were calculated without the resonant terms $if''_n$. Consequently, another explanation for the observed anomalous signals from Friedel pairs is needed; one that is not based on the differences of the square modulii of the structure factors.

One possible explanation comes from that old interpretation on the contrast of Kossel lines provided by Laue \cite{laue1949} and on Borrmann anomalous transmission \cite{borr1954, batt1962, batt1964a} were the photoabsorption is proportional to, or modulated by, the intensity of the standing wave field formed in crystals undergoing diffraction. 

However, to a general explanation of the anomalous signal based on standing waves, there is an important difference regarding the current understanding on this subject; \cite{batt1964a, truc1976, bedz1984, bedz1985, auth2001} the positioning of the standing wave field must be independent of the absorbency of the crystals \cite{more2005b}, as well as invariant regarding arbitrary choice of origin for the atomic positions. \cite{batt1964a} A detail discussion on the dependence of the standing wave field with the phase of the structure factor has been carried out by Bedzyk \& Materlik \cite{bedz1985}. But, on its usual expression the phase $\alpha_H$ of the structure factor in Eq.~(\ref{ssfeqn}) depends on absorption.

To demonstrate that absorption modulation by standing waves (AMSW) can provide a significant contribution to the anomalous signal, the incident ${\bm E}_{Inc}$ and reflected ${\bm E}_{Ref}$ waves as a function of depth $h$ are approximated to

\begin{equation}
{\bm E}_{Inc}=E_0(h)\hat{\bm v}_0\>e^{i(\omega t - {\bm k}_0\cdot{\bm r})}
\label{eieqn}
\end{equation}
and 

\begin{equation}
{\bm E}_{Ref}=R_N(\theta)E_0(h)\hat{\bm v}\>e^{i(\omega t - {\bm k}\cdot{\bm r})}
\label{ereqn}
\end{equation}
where ${\bm k}_0$ and ${\bm k}$ are the wavevectors of the waves, while $\hat{\bm v}_0$ and $\hat{\bm v}$ are their oscillation directions so that, $|C|=\hat{\bm v}\cdot\hat{\bm v}_0$ for both $\sigma$ and $\pi$ components of polarization. A comparable reflection coefficient at all depth is assumed, i.e. $R_N(\theta)$ do not depend on $h$. All dependence with depth is accounted by $E_0(h\rightarrow\infty)\rightarrow0$, which is a smooth function with very small variation over the lattice period. 

\begin{figure}
\includegraphics[width=3.2in]{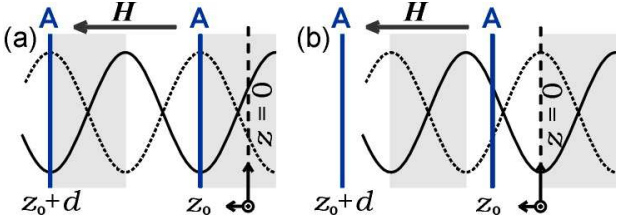}
\caption{Standing wave positioning must be invariant regarding an arbitrary choice of origin \cite{batt1964a}. Such requirement is accomplished in the present approach when the structure factor is calculated with the `` - '' signal in the phase factor. For a lattice plane made of a single atomic layer A, the structure factor phase is given by (a) $\delta_H = -2\pi z_0/d$ or (b) $\delta_H = +2\pi z_0/d$. As the crystal is rocked through reflection H, the standing waves are formed at the positions given by dark lines, move across the gray shaded areas, and vanish at the dashed-line positions. In (a) the scanned portions of the lattice planes (gray areas) are invariants regardless the value of $z_0$, but the same invariance is not observed in (b). }
\end{figure}

The time average intensity of the standing waves is then 

\begin{eqnarray}
I_{SW} &=& |\bm{E}_{Inc.}+\bm{E}_{Ref.}|^2 =\\
       &=& |E_0(h)|^2\{1+|R_N(\theta)|^2+2|R_N(\theta)||C|\cos\Phi \} \nonumber 
\label{isweqn}
\end{eqnarray}
where $\Phi=2\pi {\bm H}\cdot{\bm r}+\Psi(\theta)$, $2\pi{\bm H}={\bm k}-{\bm k}_0$, and \\ $2\pi{\bm H}\cdot{\bm r}=2\pi z/d$. The node positioning along the lattice plane normal direction $\hat{\bm z}$, which is parallel to the diffraction vector ${\bm H}=(1/d)\hat{\bm z}$, is obtained from the condition $\Phi=2\pi$, so that

\begin{equation}
z_{node}=d(1-\Psi/2\pi)=
\begin{cases} d(1/2-\delta_H/2\pi) & \text{for $\Omega=\pi$,}\\
d(1-\delta_H/2\pi) & \text{for $\Omega=0$.}
\end{cases}
\label{npeqn}
\end{equation}

Here, the positioning of the standing waves are already independent of absorption since $\delta_H$ came from the definition in Eq.~(\ref{sfeqn}); and the condition of invariance with an arbitrary choice of origin is fulfilled by the `` - '' signal of the phase factor in Eq.~(\ref{mtof}). It can be demonstrated by the following example.

For a given reflection, whose element of periodicity is made of a single atomic layer A, as shown in Fig.~5, the origin is taken arbitrarily so that it does not necessarily coincident with the atomic layer, i.e. $z_0\neq0$. When the structure factor phase is $\delta_H=-2\pi z_0/d$, according to Eq.~(\ref{npeqn}), the node position moves from $z_{node}=d/2+z_0$ ($\Omega=\pi$) to $z_{node}=d+z_0$ ($\Omega=0$) across the rocking curve, as represented by the gray shaded area in Fig.~5(a). It is an invariant behavior since the nodes will always scan the same portion (gray area) of the lattice planes regardless the value of $z_0$. On the other hand, the same invariant behavior does not occur if the geometrical phase factor is calculated with the `` + '' (plus) signal, i.e. $\delta_H=+2\pi z_0/d$, as shown in Fig.~5(b).

\begin{figure}
\includegraphics[width=3.2in]{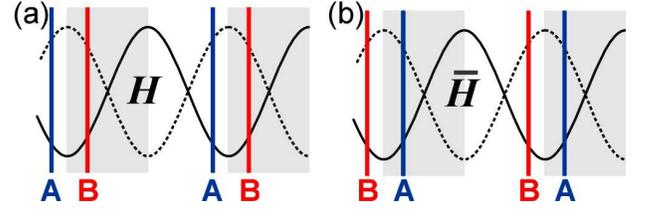}
\caption{In noncentrosymmetric crystals, ${\rm A}\neq{\rm B}$, the standing waves glide over regions (gray areas) with different electron densities during the rocking-curves of the (a) H and (b) $\bar{\rm H}$ reflections. In centrosymmetric crystals, A = B, identical regions are scanned by the nodes}
\end{figure}

One may note that the gliding sense of the nodes, in the sense of the diffracting vector in Fig.~(5), is opposite to that reported on most articles on this matter. The gliding sense is determined by the rotation sense of the dynamical phase shift $\Omega(\theta)$. A clockwise rotation sense of $\Omega$ (from $\pi$ to 0) is that predicted by the dynamical theory (e.g. Ref. \onlinecite{auth1986}), and it is the same of that obtained here for the $R_N(\theta)$ coefficient (see inset of Fig.~3). Then, the gliding sense of the nodes here should be the same of that reported by other authors. 

Assumption of an opposite gliding sense, in the sense of $\bar{H}$, would be necessary to math fluorescence results monitored during a rocking curve with the integrated intensities of a Friedel pair calculated according to ${\cal F}_H$ (with the `` + '' signal in the phase factor). \cite{truc1976}

For the invariance of the wavefield regarding the choice of origin, either a clockwise or counterclockwise rotation sense of $\Omega$ would require the `` - '' signal in the phase factor. And, to avoid any confusion regarding the gliding sense of the nodes in the remaining parts of this article, the geometrical phase factor of ${\cal F}_H$ is henceforth calculated with the `` - '' signal .

Since the nodes glide half a lattice plane distance as the crystal is rocked across the reflection domain, the portion of the lattice plane scanned by the nodes, e.g. the gray areas in Fig.~5, can be different for the H and $\bar{\rm H}$ reflections depending on the symmetry of the crystalline structure.

In centrosymmetric crystals the structure factor phase ~\textemdash~ when calculated with the `` - '' signal in the phase factor ~\textemdash~ set the nodes to scan the lattice planes from the symmetry center with lower electron density to the one with higher electron density. \cite{laue1949} It implies that, equivalent portions of the lattice planes are scanned during rocking-curves of Friedel related reflections and, consequently, the effective photoabsorptions are exactly the same for both reflections. It is illustratively shown in Fig.~6 when assuming identical A and B layers of atoms.

On the other hand, in noncentrosymmetric crystals, the scanned portions of the lattice planes for the H and $\bar{\rm H}$ reflections may not be exactly the same and then, the effective photoabsorptions for some Friedel pairs can be different. This situation can also be illustrated by the example in Fig.~6 assuming now that the A and B layers of atoms have different absorption properties. 

\begin{table} 
\caption{\label{tab1} Intensity variation of $hkl/\bar{h}\bar{k}\bar{l}$ Friedel reflections in a GaSb crystal at photon energy of 10.0 KeV. $Q_F$ and $Q_A$ stand for theoretical anomalous signals when the integrated reflectivities are taken as proportional either to $|{\cal F}_H|^2$ [Eq.~(\ref{ssfeqn}) with the `` - '' signal in the phase factor] or to ${\cal A}_H$ [Eq.~(\ref{ireqn})], respectively. $r=Q_F/Q_A$ and $r^*$ regards this ration at 10.4 KeV (above the Ga edge at 10.37KeV). Crystal dimension larger than 20$\mu m$ in all directions (see Fig.~8).}
\begin{center}
\begin{tabular}{ccccccc}
$hkl$ & ~~$Q_F$ (\%)~~ & $|{\cal F}_H/{\cal F}_{\bar H}|^2$ & ~~$Q_A$ (\%)~~ & ${\cal
A}_H/{\cal A}_{\bar H}$ & $r$ & $r^*$ \\
\hline
111 & -5.817 & 0.890 & -4.168 & 0.920 & 1.40 & 1.48 \\
113 & +6.845 & 1.147 & +4.853 & 1.102 & 1.41 & 1.52 \\
133 & -7.474 & 0.861 & -5.255 & 0.900 & 1.42 & 1.55 \\
115 & -7.906 & 0.853 & -5.519 & 0.895 & 1.43 & 1.58 \\
135 & +8.202 & 1.179 & +5.688 & 1.121 & 1.44 & 1.61 \\
117 & +8.450 & 1.184 & +5.786 & 1.123 & 1.46 & 1.67 \\
555 & -8.209 & 0.848 & -5.531 & 0.895 & 1.48 & 1.76 \\
\hline
\end{tabular}
\end{center}
\end{table}

To calculate the effective absorption coefficient $\mu_{eff}(\theta)$, as a function of the rocking angle, the normalized intensity of the standing waves

\begin{eqnarray}
\tilde{I}_{SW}(\theta, {\bm r}) &=& \frac{I_{SW}}{\frac{1}{d}\int_{0}^{d}I_{SW}(z){\rm
d}z}=\nonumber 
\\
&=& 1+\frac{2|R_N(\theta)||C|}{1+|R_N(\theta)|^2}\cos (2\pi {\bm H}\cdot{\bm r}+\Psi),
\label{nsweqn}
\end{eqnarray}
is used as a weight function \cite{batt1962} for the atomic absorption cross sections in Eq.~(\ref{laceqn}), which provides 

\begin{equation} 
\mu_{eff}(\theta) = \frac{2r_e\lambda}{V_c}\sum_n \tilde{I}_{SW}(\theta, {\bm r}_n)f_n^{\prime\prime}.
\label{amsweqn} 
\end{equation} 

By implementing the absorption probability $a~=~\mu_{eff}(\theta) d / \sin\theta_B$, the reflectivity curves show their well-known asymmetric profiles with lower intensity at the right-hand shoulder. \cite{kuri1997} 

For the sake of comparison, let define the amplitude of the anomalous signal from reflections H and $\bar{\rm H}$ by

\begin{equation}
Q_F=\frac{|{\cal F}_H|^2 - |{\cal F}_{\bar H}|^2}{|{\cal F}_H|^2 + |{\cal F}_{\bar H}|^2},
\label{qfeqn}
\end{equation}
as in structure determination methods, and 

\begin{equation}
Q_A=\frac{{\cal A}_H - {\cal A}_{\bar H}}{{\cal A}_H + {\cal A}_{\bar H}}
\label{qaeqn}
\end{equation}
where

\begin{equation}
{\cal A}_H=\int|R_N(\theta)|^2{\rm d}\theta.
\label{ireqn}
\end{equation}
$R_N(\theta)$ is given in Eq.~(\ref{refleqn}) for both reflections but calculated for different absorption probabilities as for the theoretical values in Table I and in Fig.~7.

\begin{figure}
\includegraphics[width=3.2in]{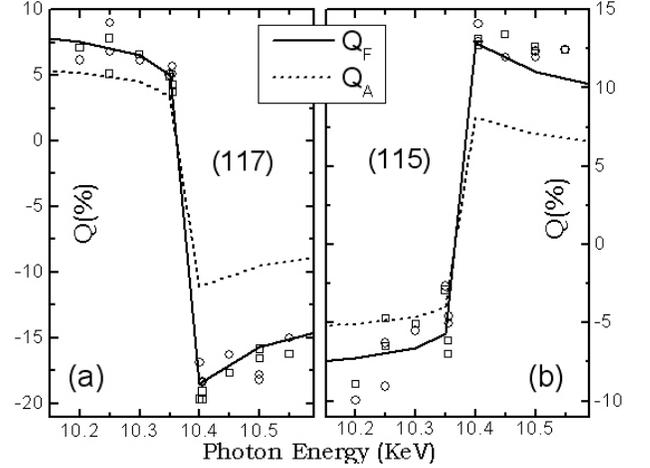}
\caption{Anomalous signals from (a) 117/$\bar{1}\bar{1}\bar{7}$ and (b) 115/$\bar{1}\bar{1}\bar{5}$ Friedel pairs in a GaSb crystal as a function of photon energy across the Ga edge at 10.37 KeV. Experimental values were determined by measuring these asymmetric reflections at either low (squares) and high (circles) incidence angles. Theoretical $Q_F$ (solid lines) and $Q_A$ (dashed lines) curves are provided by Eqs.~(\ref{qfeqn}) and (\ref{qaeqn}), respectively. Sample: commercial GaSb(001) wafer, $500 \mu$m thick, and polished both sides. X-ray source: synchrotron radiation, $\sigma$ polarization, and double-bounce Si(111) monochromator.}
\end{figure}

Since the anomalous signal in the AMSW hypothesis is related to the formation of standing waves, $Q_A$ is a function of the crystal thickness. Fig.~8 shows both the integrated reflectivity and $Q_A$ as a function of thickness for a few Friedel reflections in the GaSb crystal. It varies since a very small crystal undergoing kinematical diffraction where the intensity increases with thickness, until the dynamical diffraction regime dominated by primary extinction where, for an absorbing crystal, the intensity no longer depends on thickness.

Essentially, the observed disagreement between the $Q_A$ and $Q_F$ values in Table I, or in Fig.~7, arrives because in the former treatment interference/diffraction effects involving the $if''$ resonant amplitudes are not taken into account. Below the Ga edge only $f''_{Sb}=4.09$ has a significant value regarding $f''_{Ga}=0.52$, while above this edge they have nearly the same value, $f''_{Sb}=3.833$ and $f''_{Ga}=3.87$, and then the disagreement is more evident and beyond experimental errors. It demonstrates that the resonant amplitudes are taking part of the diffraction phenomenon.

On the other hand, the comparison in Fig.~7 also indicates that more than 50\% of the anomalous signal amplitude $Q$ relies exclusively on the effects of absorption modulation, which implies that $Q$ may depend on the crystal dimensions.

As pointed out in Fig.~8, the relative behavior of $Q$ as a function of thickness regarding strong ($Q_{strong}$) and weak ($Q_{weak}$) reflections can lead to a procedure to check the AMSW effect on experimental $Q$ values, i.e. kinematical diffraction: $Q_{strong}~>~Q_{weak}$; dynamical diffraction: $Q_{strong}~<~Q_{weak}$.

In low-absorbing crystals the primary extinction length can be of the order of hundreds of microns. Correcting the experimental anomalous-signal by a thickness related factor might be necessary to improved the accuracy of phasing methods in such small crystals.

\section{Atomic resonances and interference effects}

The incompatibility of the Stokes relation, Eq.~(\ref{bse1}), with the usage of $if''$ in the expression of the structure factor can be demonstrated by replacing Eqs.~(\ref{ssfeqn}) and (\ref{rlpeqn}) into Eq.~(\ref{bse2}), which provides 

\begin{equation}
|{\cal F}_H|\sin\alpha_H + |{\cal F}_{\bar{H}}|\sin\alpha_{\bar{H}}=0
\label{lasteqn}
\end{equation}
since $|T|$ and $|\bar{T}|$ differ from unit by less than $10^{-6}$. 

\begin{figure} 
\includegraphics[width=3.2in]{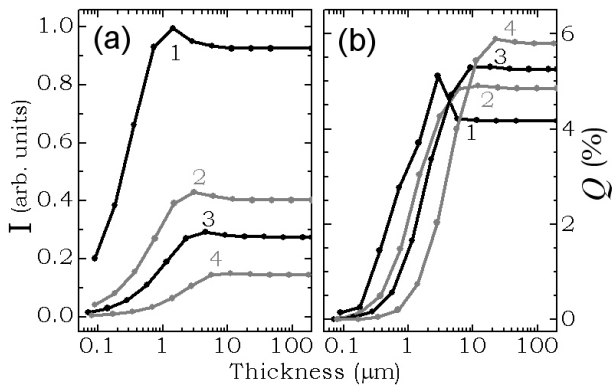} 
\caption{(a) Mean integrated reflectivity $I = C({\cal A}_H+{\cal A}_{\bar H})/2$, and (b) anomalous-signal amplitude $Q=|Q_A|$, as a function of crystal thickness. Friedel pairs: $111/{\bar 1}{\bar 1}{\bar 1}$ (1), $113/{\bar 1}{\bar 1}{\bar 3}$ (2), $133/{\bar 1}{\bar 3}{\bar 3}$ (3), and $155/{\bar 1}{\bar 5}{\bar 5}$ (4). Normalization constant $C^{-1}~=~85\mu$ radians. Crystal: GaSb. Photon energy: 10 KeV.}
\end{figure}

In noncentrosymmetric crystals, this equality is not accomplished for Friedel reflections where $|{\cal F}_H|\neq|{\cal F}_{\bar{H}}|$, as can be easily checked by just replacing a few values. It has been very difficult to understand the physical reason behind this incompatibility. Does it prove that the resonant amplitude $if''$ should not appear in the calculus of the structure factors? But then, how to explain all the evidences mentioned at the Introduction section in favor of using the complex atomic scattering factor $f$ in diffraction theories, as well as the experimental results in Fig.~7?

In the calculation of anomalous X-ray scattering factors, ``the scattering amplitude for light by a bound electron is proportional to $f$. The value of $f$ is computed as a sum over all intermediate electron states except those states occupied by other atomic electrons, and a further sum must be taken over all the electrons of the atom. When the set of intermediate states is complete, the sums provide $f_0$ the main part of the scattering factor for X-rays."

The above paragraph summarize the initial assumptions made by Cromer \& Liberman \cite{crom1970} in their paper on anomalous X-ray scattering. It clearly demonstrates that the anomalous-resonant scattering with amplitude $if''$ is a privilege of excited atoms.

\begin{figure}
\includegraphics[width=3.2in]{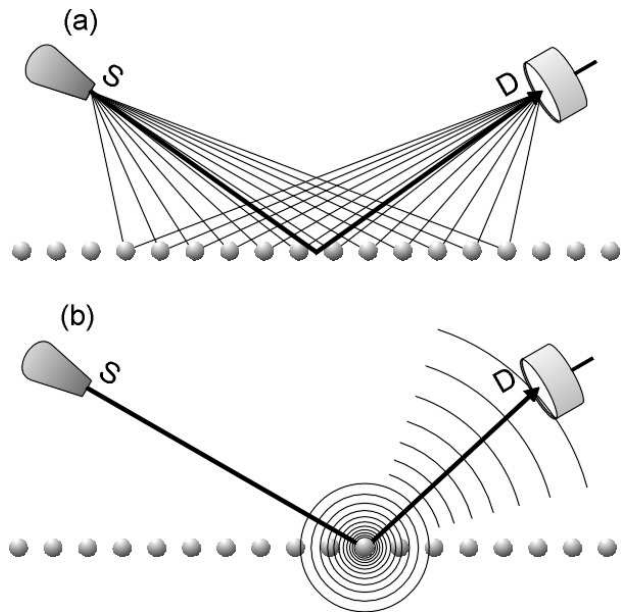}
\caption{(a) Interference of probability amplitudes of all possible photon trajectories from the source (S) to the detector (D) providing the specular reflection (dark line trajectory) by an uniform plane of atoms, see for instance Ref. \onlinecite{feyn1985}, chapter 2. (b) When an atom of the plane is excited by the incident photon, all other possible trajectories collapse into the single trajectory that goes from S to the excited atom. Then, the pattern of interference providing the specular reflection is destroyed, and any further atomic scattering, generated as a consequence of the absorbed photon energy, should not interfere with the scattering from other atoms. }
\end{figure}

Since the energy exchange between atoms and the electromagnetic fields is quantified, excited atoms inevitably imply in previous photoabsorption, which requires the photons to manifest themselves as particles. Consequently, the probability to find the absorbed photon in the entire space occupied the electric field has instantaneously collapsed at the position of the excited atom, as pictorially shown in Fig.~8(b). Since the quantum probability amplitude has been collapsed at one atom, the resonant amplitudes from this atom should not interfere with the scattering amplitudes from other atoms, unless some stimulated emission mechanisms are involved. Therefore, would not be plausible to assume that the anomalous-resonant scattering with amplitude $if''$, when occurring only as a consequence of photoabsorption at random instants of time, should not take part in interference phenomena such as X-ray diffraction? 

This is one possible explanation capable to conciliate both the Stoke relation in Eq.~(\ref{lasteqn}) and atomic resonant theories predicting complex atomic scattering factors, such as $f = f_0+f'+if''$. To be in agreement to this explanation, the $if''$ term has to be omitted in the structure factor expression, as for instance in the calculation of the specular reflection coefficients given in Eq.~(\ref{rlpeqn}) where only the atomic scattering amplitudes $f_0+f'$ take part of interference effects. \cite{stan1992} Accounting for stimulated emissions in phase with the standing wave field may be attempted if the anomalous signal dependence with the crystal dimension is experimentally confirmed.

\section{Conclusions}

By applying the Stokes relation to each lattice plane of a single crystal, the Darwin-Prins rocking curve \cite{darw1914a, darw1914b, prin1930, zach1945, kato1992} is obtained by a simple set of recursive equations valid for absorbing crystals of arbitrary thickness. However, this description is not conciliable with the usage of the anomalous-resonant term in the structure factor expression. An analogous explanation of that given by Laue \cite{laue1949, borr1954} on the contrast of the Kossel lines is necessary to explain most of the intensity difference of Friedel reflections in noncentrosymmetric crystals.

\begin{acknowledgments}
The authors would like to thank Dr. M\'arcia Fantini for valuable discussions, as well as for the very kind revision of the manuscript. This work was supported by the Brazilian founding agencies FAPESP, grant number 02/10387-5, and CNPq, proc. number 301617/95-3.
\end{acknowledgments}

\end{document}